\documentstyle[multicol,aps,prl,epsf]{revtex}
\begin{document}
\title{\bf Exact Asymptotics for One-Dimensional Diffusion with Mobile Traps}

\author{Alan J. Bray and Richard A. Blythe}

\address{Department   of   Physics   and  Astronomy,   University   of
Manchester, Manchester, M13 9PL, UK}

\date{\today}

\maketitle

\begin{abstract}
We consider a diffusing particle, with diffusion constant $D'$, moving
in one  dimension in  an infinite sea  of noninteracting  mobile traps
with  diffusion constant  $D$ and  density  $\rho$. We  show that  the
asymptotic  behavior of  the survival  probability,  $P(t)$, satisfies
$\lim_{t \to  \infty} [-\ln  P(t)]/\sqrt{\rho^2 D t}  = 4/\sqrt{\pi}$,
independent of $D'$.  The result  comes from obtaining upper and lower
bounds on  $P(t)$, and showing  that they coincide  asymptotically. We
also obtain exact results for $P(t)$ to first  order in $D'/D$  for an
arbitrary finite number of traps.

\medskip\noindent   {PACS  numbers: 05.40.-a, 02.50.Ey, 82.20.-w}

\end{abstract}

\begin{multicols}{2}
The asymptotics of the survival probability, $P(t)$, for a particle 
diffusing among mobile traps is a longstanding challenge. Over ten  
years ago Bramson and Lebowitz \cite{BL} showed by rigorous arguments 
that in one dimension $P(t)$ has a stretched exponential decay for 
large $t$, 
\begin{equation}
P(t) \sim \exp(-\lambda t^{1/2})\ .
\label{BL}
\end{equation} 
The computation of the constant $\lambda$, however, has thus far proved 
intractable, both analytically and numerically. This longstanding 
problem is resolved in this Letter. 

The problem addressed  by Bramson and Lebowitz was  first posed almost
20 years ago in the  seminal paper of Toussaint and Wilczek (TW) \cite{TW}.
These authors introduced the  two-species annihilation process, $A + B
\to 0$, as a model  of monopole-antimonopole annihilation in the early
universe,  though  applications  to  chemical kinetics  and  condensed
matter physics are more numerous \cite{Privman}. TW showed that if
the initial densities, $\rho_A(0)$ and $\rho_B(0)$, of the $A$ and $B$
particles  are equal (and  the particles  are randomly  distributed in
space),  the densities  decay  asymptotically as  $t^{-d/4}$ in  space
dimensions  $d<4$.  If  the  initial densities  are  {\em  different},
however, the  density of the  minority species ($A$, say)  decays much
more rapidly.

The connection with the trapping  problem is as follows. At late times
$\rho_A(t) \ll  \rho_B(t)$ and  the $A$ particles  can be  regarded as
independently diffusing in a background of the majority $B$ particles,
which  act  as   traps  for  the  $A$  particles   by  virtue  of  the
annihilation  reaction.   An  equivalent  problem, therefore,  is  to
consider a  single $A$ particle  moving among $B$ particles  (which do
not interact  with each  other) and ask  for the  probability, $P(t)$,
that the  $A$ particle survives  to time $t$.   In the context  of the
original $A + B \to 0$ process  (or $A + B \to B$ \cite{OZ}, which has
the same asymptotics \cite{BL}) the ``particle'' $A$ and ``traps'' $B$
are taken to have the  same diffusion constants, but for generality we
will take them to be different, $D'$ and $D$ respectively.

To our knowledge, no analytical result has been obtained up to now for
the constant  $\lambda$ in Eq.\ (\ref{BL}).   Furthermore, attempts to
determine $\lambda$  by numerical simulations (or even  to confirm the
predicted  stretched-exponential   decay  with  exponent   $1/2$)  are
severely    hampered   by    large,    slowly   decaying    transients
\cite{SSB1,SSB2,MG,OB}.  In ref.\ \cite{MG}, a sophisticated numerical
approach enabled data to be obtained down to $P(t) \sim 10^{-35}$, but
still  the  asymptotic  time  dependence could  not  be  unambiguously
established.

For  this trapping  problem, the  traps  are infinite  in number,  and
distributed  randomly  on  the  interval  ($-\infty$,  $\infty$)  with
density $\rho$. By  contrast, if the number of  traps is {\em finite},
the problem  is equivalent to  the much studied  predator-prey problem
\cite{Sid1,Sid2}, where  the traps  are predators and the particle  is 
the  prey. In  this case  the prey  survival probability decays as a 
power law,
\begin{equation}
P(t) \sim t^{-\theta(N_L,N_R,D'/D)}\ ,
\label{theta}
\end{equation}
where $N_L$ ($N_R$)  is the number of predators  initially to the left
(right) of the prey. To our knowledge the only exactly-solved examples
are for $N_L + N_R \le  2$. Obtaining analytical results for more than
two predators is another longstanding challenge.

In  this Letter  we obtain  two  exact analytical  results. First,  we
finally resolve the  question of  the asymptotics of  $P(t)$ for  the 
trapping problem.   We verify  the  asymptotic form  (\ref{BL}), and  
determine exactly  the  value  of  the  constant $\lambda$,  namely  
$\lambda  = 4\rho(D/\pi)^{1/2}$, i.e.\
\begin{equation}
P(t) \sim \exp[-4\rho(Dt/\pi)^{1/2}]\ .
\label{exact}
\end{equation}
Note that this asymptotic result  depends only on the density, $\rho$,
and diffusion constant,  $D$, of the traps and  is {\em independent of
the  diffusion constant  $D'$ of  the  particle}. The  value of  $D'$,
however,  does  affect  the  rate  of approach  to  asymptopia.   Eq.\
(\ref{exact})  is derived  by  obtaining upper  and  lower bounds  for
$P(t)$, and showing that they converge for large $t$. Furthermore, the
form  of  the  lower  bound  suggests why  large  corrections  to  the
asymptotic behavior might be expected.

Secondly, we outline \cite{BB1} a perturbation theory in $D'/D$, based
on  a  path-integral  formulation, for the computation of the exponent 
$\theta$ in Eq.\ (\ref{theta}). For small $D'/D$ and arbitrary values 
$N_L$ and $N_R$, we find
\begin{equation}
\theta =  \frac{N}{2} + \frac{1}{\pi}\,(N  - \Delta^2)\,\frac{D'}{D} +
O\left(\frac{D'^2}{D^2}\right)\ ,
\label{pert}
\end{equation}
where $N=N_L+N_R$ is the total number of predators and $\Delta = N_L -
N_R$ measures their left-right asymmetry with respect to the prey.

We  will first  describe the  treatment of  the trapping  problem with
infinitely many traps.  The bounds  that lead to Eq. (\ref{exact}) are
obtained as follows. 

\noindent\underline{\bf  1.  Upper  Bound.}  An  obvious  upper bound, 
$P_U(t)$, on  $P(t)$, for  any $D'$, is  provided by the  problem with 
$D'=0$, in which the particle  stays at its initial position, which we 
call $x=0$.   Although we have as  yet been unable to  make this bound
rigorous \cite{BB},  it is intuitively  clear that when, as  here, the
traps  are (statistically)  symmetrically placed  with respect  to the
particle,  the particle  will on  average survive  longer if  it stays
still  than  if  it  diffuses. This assertion has been checked, using 
the algorithm outlined in ref.\cite{MG}, for all (lattice) walks up to 
time $t=28$. It is also supported  by  Eq.\ (\ref{pert}), which  shows 
that  for any symmetric  case ($\Delta=0$), the decay of $P(t)$ is 
faster for small $D'$ than for $D'=0$, for any {\em finite} $N$.

For $D'=0$ (sometimes called the ``scavenger model'' \cite{RK} in  
predator-prey terminology),  $P(t)$ is  just the  probability that
none of the moving traps has  reached the origin up to time $t$. This
problem is exactly soluble  \cite{BZK}. Since similar techniques will 
be needed to  derive the  lower bound,  we outline  the solution here.

The $N$ traps move independently according to the Langevin
equations
\begin{equation}
\dot{x_i} = \eta_i(t)\ ,\ \ i=1,\ldots,N\ ,
\label{Langevin}
\end{equation}
where  $\eta_i(t)$  is  Gaussian   white  noise  with  mean  zero  and
correlator
\begin{equation}
\langle \eta_i(t)\eta_j(t') \rangle = 2D\, \delta_{ij}\delta(t-t')\ .
\label{correlator}
\end{equation}
The  quantity  $P(t)$ is  just  the  product  of the  individual  trap
probabilities.   For a  given  trap starting  at  $x_i$, the  required
probability      is      \cite{Sid2}      $P_1(x_i,t)      =      {\rm
erf}(|x_i|/\sqrt{4Dt})$. So our upper bound is
\begin{equation}
P_U(t)=\left\langle\prod_{i=1}^N{\rm erf}\left(|x_i|/\sqrt{4Dt}\right)
\right\rangle\ ,
\end{equation}
where  $\langle \ldots  \rangle$  means an  average  over the  initial
positions  of the  traps. Since  the $x_i$  are also  independent, the
latter average also factors. Using $N=\rho L$, where $L$ is the length
of   the  system,  and   each  $x_i$   is  uniformly   distributed  in
$(-L/2,L/2)$, gives
\begin{eqnarray}
P_U(t)   &   =  &   \left[1   -   1/L   \int_{-L/2}^{L/2}  dx\,   {\rm
erfc}(|x|/\sqrt{4Dt})\right]^{\rho   L}    \nonumber   \\   & \to  &
\exp[-4\rho(Dt/\pi)^{1/2}]\ ,
\label{upper}
\end{eqnarray}
where the final result follows on taking the limit $L \to \infty$.

\noindent\underline{\bf 2. Lower Bound.}  Consider the same system as
before but  with a  pair of absorbing  boundaries at $x=\pm  l/2$.  We
consider the subset  of initial conditions in which  all the traps lie
outside the interval $(-l/2,l/2)$ (and  the particle is at $x=0$), and
trajectories in  which neither  the particle nor any of the  traps has
crossed  a boundary  up to  time  $t$. We  calculate the  probability,
$P_L(t)$,  of such  an occurrence  over  the ensemble  of all  initial
conditions and trajectories.   These restricted initial conditions and
trajectories are a  subset of all the possible  initial conditions and
trajectories in which the particle never meets a trap. It follows that
$P(t) \ge P_L(t)$, i.e.\ $P_L(t)$ is a lower bound.

The probability that  there are no traps in  the interval $(-l/2,l/2)$
at $t=0$  is $\exp(-\rho l)$.  Given  that there are no  traps in this
interval at $t=0$, the probability that no traps enter the interval up
to time $t$  is given by the  same result as in the  derivation of the
lower   bound,  namely  $\exp[-4\rho(Dt/\pi)^{1/2}]$.    Finally,  the
probability that  the particle,  starting at $x=0$,  has not  left the
interval  $(-l/2,l/2)$ up  to  time $t$  is  given, for  times $t  \gg
l^2/D'$, by \cite{Sid2} $(4/\pi)\exp(-\pi^2D't/l^2)$. Assembling these
contributions gives
\begin{equation}
P(t) \ge (4/\pi)\exp[-4\rho(Dt/\pi)^{1/2}-(\rho l + \pi^2D't/l^2)]\ .
\label{lower}
\end{equation}

Since  this inequality  holds for  all $l$,  the best  lower  bound is
obtained by maximizing  with respect to $l$. The  optimum value is 
$l=(2\pi^2D't/\rho)^{1/3}$, and the best lower bound is
\begin{equation}
P_L(t)=\frac{4}{\pi}\exp[-4\rho(Dt/\pi)^{1/2}-3(\pi^2\rho^2 D't/4)^{1/3}]\ .
\label{lower1}
\end{equation} 
Since the  second term in the  exponent is negligible  compared to 
the first as  $t \to  \infty$, the two  bounds converge to yield the 
asymptotic form,  Eq.\ (\ref{exact}), for $P(t)$. More precisely, we 
can take the logarithm of $P(t)$ and divide out the leading $\sqrt{t}$ 
dependence to get 
\begin{equation}
\frac{4}{\sqrt{\pi}}\le    -\frac{\ln    P(t)}{(\rho^2Dt)^{1/2}}   \le
\frac{4}{\sqrt{\pi}}        +        3\left(\frac{\pi}{2}\right)^{2/3}
\frac{(D'/D)^{1/3}}{(\rho^2 Dt)^{1/6}}\ ,
\label{inequalities}
\end{equation}
giving  $\lim_{t \to  \infty}  -[\ln  P(t)]/(\rho^2 Dt)^{1/2} 
= 4/\sqrt{\pi}$.

As  an  aside  we  note   that,  while  the  left-hand  inequality  in
(\ref{inequalities}) holds for {\em all} $t$, since Eq.\ (\ref{upper})
does, the right-hand  inequality is strictly a large  $t$ result. This
is   because   the   factor   $(4/\pi)\exp(-\pi^2D't/l^2)$   in   Eq.\
(\ref{lower}) comes from the  lowest mode in the Fourier decomposition
of  the   survival  probability  of  the  particle   in  the  interval
$(-l/2,l/2)$.   This mode dominates  for $D't  \gg l^2$, which requires 
$\rho^2 D't \gg 1$. A lower  bound on $P(t)$ valid for all $t$  can 
easily be written down  by   including  all  Fourier  modes,  but  the  
large-$t$  form (\ref{lower1}) is sufficient for present purposes.

In Figure 1, the left and right hand sides of Eq.\ (12), representing 
the two asymptotic bounds, are plotted and compared to the numerical 
data of ref.\cite{MG}. The data are plotted using the dimensionless time 
$\rho^2Dt$, and the axes are chosen to test the asymptotic form 
(\ref{exact}). The data lie between the bounds (except at early times where 
lattice effects are important \cite{OB}). The bounds displayed are for the 
case $D'=D$. The very slow convergence of the bounds is consistent with the 
observed trend in the data.   

\begin{figure}
\narrowtext
\centerline{\epsfxsize=\columnwidth\epsfbox{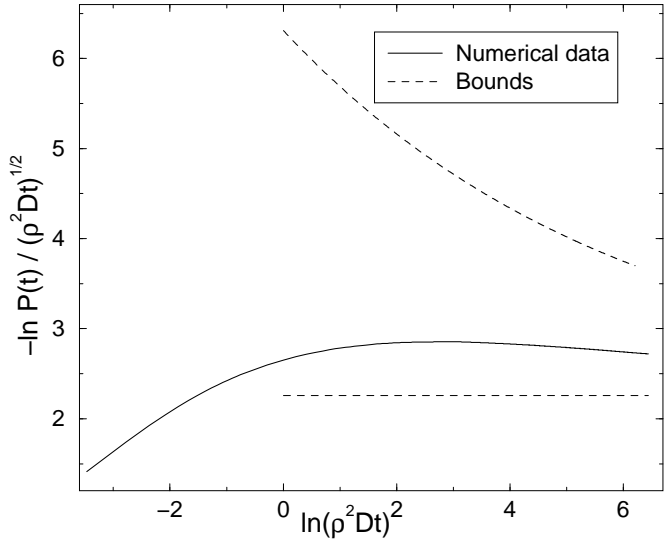}}
\caption{Numerical data from ref.[7] and the upper and lower bounds,  
Eq.\ (12), for $D'=D$. The numerical data were generated with $D=D'=1/2$ 
and $\rho=1/4$.} 
\end{figure}

Some comments  on Eq.\ (\ref{inequalities}) are in  order.  First, the
right-hand  inequality  already  suggests  the  possibility  that  the
approach  to asymptopia  could  be  {\em very}  slow,  as observed  in
simulations, since  the second term  vanishes only as  $t^{-1/6}$. For
the  case $D'=D$ that is usually simulated, the asymptotic  result
(\ref{exact}) gives, for a dimensionless time $\rho^2Dt =  10^4$ say, 
a survival probability of order $10^{-98}$. This is far smaller than 
can be reached in simulations (or experiments!) even with sophisticated 
methods \cite{MG}. Yet even at this large time, the  second  term on  
the  right-hand  side  of (\ref{inequalities}) is  still 39\%  as big 
as  the first! This term would give a further factor of $10^{-38}$ in 
the survival probability, i.e.\ a total probability of order $10^{-136}$! 
Of course, this  second term is only a bound,  and the true correction
to the  asymptotic form  could be  smaller. The form  of the  bound is
nonetheless suggestive and illustrates how large subdominant terms could 
arise. A clearer understanding of these subdominant terms is necessary  
before a detailed comparison with numerical or experimental data can be 
made. 

A second interesting point is that the mean-square displacement of the
particle, averaged over the  surviving trajectories used to derive the
lower bound, grows as $l^2 \sim (D't/\rho)^\nu$ with $\nu=2/3$. This
can be  compared with the  estimate $\nu =  0.5$ to 0.6  obtained from
recent simulations  \cite{MG}.  Note, however,  that these simulations
were not in  the asymptotic regime, so the  numerical estimates should
be  treated with  caution.  Furthermore,  the value  $\nu=2/3$ obtained
from the  lower bound  for $P(t)$ does  not necessarily  represent any
kind of bound for $\nu$. Nevertheless, the rough agreement between the
measured value  and that  obtained from our  very simple  arguments is
again suggestive.

The model can be generalized to include $n$ diffusing particles starting 
from the same point \cite{MG}. Our bounds can easily be generalized 
to this case \cite{BB1}. The probability that all $n$ particles survive 
at time $t$ satisfies the same inequalities (\ref{inequalities}) except 
that the final term on the right-hand side acquires an additional factor 
$n^{1/3}$. So the bounds converge for $t \to \infty$ and the same 
asymptotic form, Eq.\ (\ref{exact}), is obtained for all $n$.

Our approach is also readily generalized to dimensions $d>1$ \cite{BB1}. 
For $d<2$ we find $P(t) \sim \exp[-a_d\rho(Dt)^{d/2}]$, where  
$a_d = (2/\pi d)(4\pi)^{d/2}\sin(\pi d/2)$ while, for $d=2$,  
$P(t) \sim \exp[-4\pi\rho D t/\ln t]$. The latter agrees with the 
functional form obtained in \cite{BL}, but with a precise value for 
the constant. For $d>2$, simple exponential decay is obtained, in 
agreement with \cite{BL}, but the bounds no longer converge  
and the decay constant cannot be determined. 

In the remainder of this Letter we sketch \cite{BB1} the derivation of 
the perturbation theory, for a finite  number of traps, that leads to 
Eq.\ (\ref{pert}).  We recall that the $N$ traps move independently  
according to Langevin  equations (\ref{Langevin}) with noise correlator 
(\ref{correlator}). We call the particle coordinate $x_0(t)$.  It obeys 
the Langevin equation $\dot{x}_0 = \eta_0(t)$, where $\eta_0(t)$ is 
independent of the other noise terms, with correlator
$\langle \eta_0(t)\eta_0(t')\rangle = 2D'\delta(t-t)$, corresponding to 
a particle diffusion constant $D'$. It is convenient to introduce the 
relative  coordinates $y_i   =   x_i   -   x_0$, $i=1,\ldots,N$. The 
corresponding  Langevin equations are $\dot{y}_i = \xi_i(t)$  where the 
noise  terms, $\xi_i(t)  = \eta_i(t)  - \eta_0(t)$, have correlators
\begin{equation}
\langle \xi_i(t)\xi_j(t') \rangle = 2D_{ij} \delta(t-t')\ ,
\end{equation}
where $D_{ij}  = D\delta_{ij}  + D'$. The  equivalent Fokker-Planck
equation, $\partial P/\partial t = \sum_{{i,j}=1}^N D_{ij}\,\partial^2
P/\partial y_i  \partial y_j$, has  to be solved subject  to absorbing
boundary conditions  at $y_i=0$ for  any $i$ (i.e.\ when  any trap 
meets  the  particle). In  principle  this equation  can  be  solved by  
a coordinate  transformation  that diagonalizes  the matrix $D$ to give 
isotropic  diffusion.  This  transformation, however,  also rotates the 
edges  of the absorbing region so that  they are no longer mutually 
orthogonal but form the edges of an $N$-dimensional hyperwedge. 
For $N=2$ this becomes a two-dimensional wedge, and  the problem  can  be 
solved exactly \cite{Sid1,Sid2}, but to our knowledge there are no exact 
solutions for $N \ge 3$.

An alternative starting point is the path-integral representation for 
the survival probability of the particle, 
\begin{equation}
P(t)     =     \frac{\int_{R}D\vec{y}(t)\,\exp(-S[\vec{y}])}     {\int
D\vec{y}(t)\,\exp(-S[\vec{y}])}\ ,
\end{equation} 
where $\vec{y}$  is a shorthand for  $(y_1,\ldots,y_n)$, $S[\vec{y}] =
(1/4)\sum_{{i,j}=1}^N (D^{-1})_{ij}\int_0^t dt'\,
\dot{y}_i(t')\dot{y}_j(t')$, and the  subscript $R$ indicates that the
path integral is  restricted to ``surviving'' paths, in  which none of
the  $y_i$ has changed  sign up to time  $t$. The  matrix $D$  is easily
inverted to give $(D^{-1})_{ij} = (1/D)(\delta_{ij} - \lambda)$, where
$\lambda = D'/(D+ND')$ will be our expansion parameter. Thus
\begin{eqnarray} 
S[\vec{y}] &  = &  S_0[\vec{y}] - S_1[\vec{y}]  \\ S_0[\vec{y}] &  = &
\frac{1-\lambda}{4D}\sum_i   \int_0^t   dt'\,   [\dot{y}_i(t')]^2   \\
S_1[\vec{y}]  &   =  &  \frac{\lambda}{2D}\sum_{i<j}   \int_0^t  dt'\,
\dot{y}_i(t')\,\dot{y}_j(t')\ ,
\end{eqnarray}
where the diagonal terms, $i=j$, have been absorbed into $S_0$.
 
A convenient  normalization of  the path integral  is provided  by the
path integral for the survival probability, $P_0(t) \sim t^{-N/2}$, of
the same problem with $D'= 0$ and $D \to D_0 = D/(1-\lambda)$. Then
\begin{equation}
\frac{P(t)}{P_0(t)}  =   \frac{\langle  \exp(S_1[\vec{y}])  \rangle_R}
{\langle \exp(S_1[\vec{y}]) \rangle}\ ,
\end{equation}
where  the   average  in  both   cases  is  over  paths   weighted  by
$\exp(-S_0[\vec{y}])$.

The  final step  is  to  expand numerator  and  denominator using  the
cumulant expansion. To first order  in $\lambda$ (i.e.\ first order in
$S_1$) one has
\begin{equation}
P(t)/P_0(t) = \exp(\langle S_1 \rangle_R - \langle S_1 \rangle)\ .
\label{P(t)} 
\end{equation} 
Since the different values of $i$ decouple in $S_0$ we have
\begin{equation}
\langle S_1 \rangle_R  = (\lambda/2D) \sum_{i<j} \int_0^t dt'\,\langle
\dot{y}_i(t')\rangle_R \langle \dot{y}_j(t')\rangle_R\ .
\label{S1av}
\end{equation} 
The  corresponding  expression for  the  unrestricted paths  vanishes,
since $\langle  \dot{y}_i(t') \rangle = 0$ by  symmetry.  The quantity
$\langle \dot{y}_i(t')\rangle_R$ is independent  of $i$ apart from the
sign: $\langle  \dot{y}_i(t')\rangle_R  > 0$  ($<0$) for traps that  
stay to  the  right (left)  of  the particle.   The calculation  of
$\langle  \dot{y}_i(t')\rangle_R$ is  straightforward. We
only require the result in the regime $t_0 \ll t' \ll t$, where $t_0 =
y_{i0}^2/D$ and $y_{i0}$ is the initial value of trap $i$. In this
regime one finds (for traps which start on the right of the particle) 
\cite{BB1}
\begin{equation}
\langle \dot{y}_i(t') \rangle_R = 2  (D_0/\pi t')^{1/2}\ ,\ t_0 \ll t'
\ll t\ .
\label{ydot}
\end{equation}
We now insert this result into Eq.\ (\ref{S1av}). 
Let there  be  $N_L$ ($N_R$) traps to the  left (right) 
of  the particle. Then  the $[N_L(N_L-1)/2 +  N_R(N_R-1)/2]$ pairs
$(i,j)$ of traps  whose members are on the {\em  same} side of the
particle  contribute  with  positive  weight  to  (\ref{S1av})  while  
the $N_LN_R$ pairs whose members are on {\em opposite} sides enter with 
negative weight. Defining $\Delta = N_L-N_R$ (and $N=N_L+N_R$) this 
gives
\begin{equation}
\langle S_1 \rangle_R = \frac{\lambda D_0}{\pi D}\,(\Delta^2 - N)\,\ln
t
\label{S1avfinal}
\end{equation}
to leading  logarithmic accuracy  for large $t$.   The factor  $\ln t$
comes from the integral $\int_{t_0}^t (dt'/t')$ that appears when Eq.\
(\ref{ydot}) is  substituted into  Eq.\ (\ref{S1av}). Using the
form (\ref{ydot}) for all $t'$, with  $t$ and $t_0$ as upper and lower
cut-offs, is correct  to leading  logarithmic accuracy \cite{BB1}. Note
that the  contributions from  the lower cut-off,  which depend  on the
initial positions of  the traps, do not contribute  to the leading
logarithm.   Using (\ref{S1avfinal})  in (\ref{P(t)}),  recalling that
$\langle S_1 \rangle = 0$ and $P_0(t) \sim t^{-N/2}$, and noting that,
to leading  order in $D'$, $\lambda =  D'/D$ and $D_0 =  D$, one finds
$P(t) \sim t^{-\theta}$ with  $\theta$ given by Eq.\ (\ref{pert}). The
power-law  decays hold  when $t  \gg  y_{0i}^2/D$ for  all $i$.   The
magnitudes  of the  initial  coordinates $y_{0i}$  determine the 
{\em amplitude} of  the power-law, while their signs determine the
{\em exponent} through the value of $\Delta$. 

Eq.\ (3)  can be  compared with the  exactly solved  cases $N_L=N_R=1$
(``surrounded  prey'') and $N_L=0,  N_R=2$ (``chased  prey'') 
\cite{Sid1,Sid2}, for  which   $\theta  =  \pi/2\alpha$  and   
$[\pi/2(\pi  -  \alpha)]$ respectively, where $\alpha  = 
\cos^{-1}[D'/(D+D')]$.  Expanding these to first order in $D'$ gives  
agreement with Eq.\ (\ref{pert}).

In summary, we have obtained the exact asymptotic form of the survival 
probability of a particle moving in an infinite background of mobile 
traps. The asymptotic form is independent of the diffusion constant of 
the particle. A perturbation expansion in the ratio of the particle and 
trap diffusion constants has been developed for the case of a finite 
number of traps.

We thank V. Mehra and P. Grassberger for supplying their raw data, and 
for a helpful correspondence. This work was supported by EPSRC grant 
GR/R53197.

\end{multicols}


\begin{references}
\bibitem{BB1}
Full details will be presented elsewhere.
\bibitem{BB} A rigorous, but weak, upper bound, 
$P_L(t) \sim \exp[-2\rho(Dt/\pi)^{1/2}]$, can be obtained by requiring 
that all the traps which were initially to the left (right) of the particle 
are to the left (right) of the particle at time $t$ (A. J. Bray and 
R. A. Blythe, unpublished).
\end{references}
\end{document}